\documentstyle[12pt]{article}

\begin{document}

\centerline{ S.Novikov\footnote{Landau Institute for Theoretical
Physics, Kosygina str. 2, Moscow 117940, Russia and IPST,
University of Maryland-College Park, MD, 20742-2431, USA, e-mail
novikov@ipst.umd.edu; the first version of this work  was
completed during the stay in Korean Institute for the Advanced
Studies (KIAS), Seoul, S.Korea in November 2002; it was partially
supported by the NSF Grant DMS-0072700. The author made several
improvements in February 2004, in particular, concerning the
reconstruction of the discrete connection for $n\geq3$. }}

\vspace{0.3cm}

\centerline{\Large Discrete Connections on the Triangulated
Manifolds }

 \centerline{\Large and  difference linear equations}

\vspace{0.3cm}

{\bf Abstract:} {\it Following the authors works \cite{N,ND1,ND2},
we develop a theory of the discrete analogs of the
differential-geometrical $GL_n$-connections  over the triangulated
$n$-manifolds. We study a nonstandard discretization based on the
interpretation of DG Connection as  the linear first order
(''triangle'') difference equation in the simplicial complexes
acting on the   scalar functions of  vertices. This theory
appeared as a by-product of  the new type of discretization for
the special Completely Integrable Systems, such as the famous 2D
Toda Lattice and corresponding 2D stationary Schrodinger
operators. A nonstandard discretization of the 2D Complex Analysis
based on these ideas   was developed in the recent work
\cite{ND3}. A complete classification theory is constructed here
for the Discrete DG Connections based on the mixture of the
abelian and nonabelian features. }

\vspace{0.2cm}

{\bf I.General Definitions: The Discrete DG Connections}.

\vspace{0.2cm}

 Let $M$ be a $n$-dimensional simplicial complex.

 By the {\bf Discrete Differentially-Geometrical (DG)
Connection} we call any set of coefficients $0\neq b_{T:P}\in
k^*,k=R,C,$ assigning a nonzero number to every pair consisting of
the $n$-simplex $T$ and  its vertex $P\in T$.

Every DG-connection defines a first order difference {\bf Triangle
Operator} $Q$ mapping
 the space of  $k$-valued functions of vertices $\psi_P$ into the space
of functions of $n$-simplices:

$$(Q\psi )_T=\sum_{P\in T}b_{T:P}\psi_P$$

Such operators played an important role in the works
\cite{N,ND1,ND2}. For the needs of the theory of discrete DG
Connections only the linear {\bf Triangle Equation} is important

$$Q\psi=0$$
 well defined up to the {\bf Abelian Gauge Transformations}
$$Q\rightarrow f_TQg^{-1}_P,\psi_P\rightarrow g_P\psi_P$$ where
$f\neq 0,g\neq 0$. Beginning from now we denote vertices by the
letters $i,j,l,...$.

Therefore for every $n$-simplex $T$ with vertices $i,j\in
(i_0,\ldots,i_n)$ only the  ratios are essential
$\mu^T_{ij}=b_{T:i}/b_{T:j}$ where $b_{T:i}$ are the coefficients
of DG connection associated to the vertices of the simplex $T$. We
assume beginning from now that {\bf the DG Connection is given by
the set of nonzero numbers $\mu_{ij}^T$ for all $n$-simplices and
pairs of their vertices.} Obviously we have
$$\mu_{ii}^T=1,\mu^T_{ij}\mu^T_{ji}=1,\mu^T_{ij}\mu^T_{jk}\mu^T_{ki}=-1$$.

Let $T,T'$ be a pair of $n$-simplices such that
 $i,j\in T\bigcap T'$. We define a {\bf Gauge-Invariant Coefficients}
$$\mu^T_{ij}\mu^{T'}_{ji}=\rho_{ij}^{TT'}$$

\newtheorem{ex}{Example}

\newtheorem{lem}{Lemma}

\begin{lem} The whole set of the gauge invariant coefficients $\rho_{ij}^{TT'}$
can be recovered from the {\bf Minimal Subset} such that $T$ and
$T'$ are the closest neighbors,  i.e. the intersections $T\bigcap
T'$ are the $(n-1)$-dimensional faces. There are ''trivial'' sets
A and B of relations on these quantities:

A.For every triangle $[ijl]\subset T\bigcap T'$ we have
$$\rho_{ij}^{TT'}\rho_{jl}^{TT'}\rho_{li}^{TT'}=1$$

B.For every closed path $T_0T_1\ldots T_NT_0$ in the Poincare dual
cell subdivision of the triangulated manifold $M$ where
n-simplices $T$ define the vertices, all pairs $T_p\bigcap
T_{p+1}$ define the edges (they are dual to the $n-1$-faces), and
the edge $<ij>$ belongs to all $T_p$,  we have
$$\prod_{p=1}^N\rho_{ij}^{T_pT_{p+1}}=1$$
\end{lem}

Proof. For the simplicial manifold $M$ every pair of $n$-simplices
$T,T'$ such that $i,j\in T\bigcap T'$ can be joined by ''path''
$T_0=T,T_1,\ldots,T_m=T'$ where $i,j\in T_k\bigcap T_{k+1}$ for
all $k=0,\ldots,m-1$, and $T_k\bigcap T_{k+1}$ are
$(n-1)$-simplices for all values of $k$.  We have by definition
$$\rho_{ij}^{TT'}=\prod_{k=0}^{k=m-1}\rho_{ij}^{T_kT_{k+1}}$$ Our
trivial set A of the relations for these quantities follows from
the same relations for the quantities $\mu^T_{ij},\mu^{T'}_{ij}$
as above. In order to prove the set of relations B we point out
that any such closed path in the dual cell decomposition can be
obtained as a product of elementary paths $T_0\ldots T_m$
corresponding to the simplicial stars of every $n-2$-simplices
$<ij>\subset\sigma\subset St(ij)$. We have here
$$St(\sigma)=T_0\bigcup\ldots\bigcup T_m$$ and the relation
$$\rho_{ij}^{T_0T_1}\ldots
\rho_{ij}^{T_mT_0}=\mu_{ij}^{T_0}\mu_{ji}^{T_1}\mu_{ij}^{T_1}\ldots
\mu_{ji}^{T_0}=0$$

 Lemma
is proved.

{\bf Problem}: Is it possible to recover the whole DG connection
from the Minimal Subset of the abelian gauge-invariant
coefficients $\rho_{ij}^{TT'}$?  Which invariants of DG connection
should be added if it is impossible?

We are going to solve this problem for the 2D and 3D  manifolds
$n=2,3$ where the whole set of the additional invariants is easy
to find out: let us choose any set of the closed combinatorial
''framed'' paths $\gamma_1, \ldots,
\gamma_{b_1},a_1,\ldots,a_{tor_1}$ representing the basis of the
homology group $H_1(M,Z)$. We define the following abelian
gauge-invariant {\bf Topological quantities}:
$$\mu(\gamma)=\prod_{\gamma}\mu_{l,l+1}^{T_l}$$ for every closed
''framed path'' $\gamma_k$   consisting of edges through the
vertices $0,1,\ldots,l,l+1,\ldots, m_k=0$, equipped by such
triangles (n-simplices) $T_l$ that $[l,l+1]\subset T_l$.

We are going to prove below the following

\newtheorem{th}{Theorem}
\begin{th}For any $n\geq 2$  the set of invariants $$\{\rho_{ij}^{TT'},
\mu(\gamma_k),\mu(a_s)\}$$
 is complete
where $i\neq j,\rho_{ij}^{TT'}, [ij]\subset T\bigcap T'$.

 For the compact oriented 2-manifolds the only nontrivial relation on
these quantities is  $$\prod_{[ij]\in M
}\rho_{ij}^{TT'}=1,\partial T=[ij]+...,\partial T'=[ji]+...,
T\bigcap T'=[ij]$$
 Here
$[ij]$ means all edges in the manifold $M$, the 2-simplices $T,T'$
are oriented as prescribed by the global orientation.

For the compact oriented n-manifolds the complete set of relations
on these quantities can be described in the following way: there
are trivial relations A for every 2-simplex $[ijl]=\Delta\subset
T\bigcap T'$ where
$$\rho_{ij}^{TT'}\rho_{jl}^{TT'}\rho_{li}^{TT'}=1$$ and the
relations B for every closed path in the Poincare dual cell
decomposition corresponding to the boundaries of the dual 2-cells.
For the description of the nontrivial relations we choose the set
of integral 2-chains $$z_1,...,z_{b_2},u_1,...,u_{tor_1}\}$$ where
 $z_1,...,z_{b_2}$ is the basis of the group $H_2(M,Z)$ and $u_s$
represent the basis of cycles mod $m_s$, i.e. $\partial
u_s=m_sa_s$. Let all chains $z,u$ be presented as sums $\sum
\Delta_k$ with ''framing'' $T_k$ for every oriented 2-simplex
$\Delta_k$, $\partial T_k=\Delta_k+...$. The nontrivial relations
have the following form: The product
$$\prod_{[ij]=\Delta_k\bigcap\Delta_{k'}}\rho^{T_kT_{k'}}_{ij}$$
is equal to one for the cycles $z$; it is  equal to $\mu(\partial
u_s)^*=\mu(a_s)^{m_s}$ for the torsion part $u_s$.

For $n=3$ exactly one global relation between these  relations
follows from the following identity: $$\prod_{[ij]\in
M}\rho_{ij}^{TT'}=1$$ in the compact oriented 3-manifold $M$ where
orientation in the 3-simplices $T,T'$ is induced by the one in
$M$,  $\partial T=[ijl],\partial T'=[jil]$.

For $n\geq 2$ any set of quantities $$\rho_{ij}^{         TT'},\mu
(\gamma_k),\mu(a_s),k=1,...,b_1,s=1,...,torsion_1$$ satisfying to
this set of trivial and nontrivial relations can be realized by
the discrete DG Connection uniquely up to the abelian gauge
transformation.

\end{th}

 \vspace{0.3cm}

 {\bf II. The Nonabelian Curvature.}

\vspace{0.2cm}

By definition, {\bf the Nonabelian Curvature} for the discrete
$GL_n$  Connection of the type described above   is the
obstruction to the existence of full $n$-dimensional space of the
local solutions to the triangle equation $Q\psi=0$ in the whole
simplicial stars of the vertices. However, on the manifold $M$ it
is enough to consider only the obstructions to the existence of
local solutions in {\bf the simplicial stars $St(\sigma)$ for all
$(n-2)$-simplices} $\sigma=[0,1,\ldots,n-2]$. The whole set of
vertices in this star contains also the complementary set
$p,p=1,2,\ldots,m$ where the number $m$ depends on $\sigma$. The
$n$-simplices in this star are exactly the following
$$T_p=[\sigma,p-1,p]$$ where $p$ is counted here modulo $m$. Every
$n$-simplex $T_p$ contains in its boundary a pair of faces inside
of this star: $$[\sigma,p-1]\bigcup [\sigma,p]\subset
\partial T_p $$

We start with initial data taking $\psi_0,...,\psi_{n-2},\psi_p$
as an arbitrary $n$-vector $\eta$. From the equation $Q\psi=0$ in
the simplex $T_{p+1}$ we obtain the value
$$\psi_{p+1}=\sum_{q=0}^{q=n-2}\mu_{q,p+1}^{T_{p+1}}\psi_q+
\mu_{p,p+1}^{T_{p+1}}\psi_p,q\in\sigma,p=1,...,m           $$

$$\psi_q=\psi_q,q=0,...,n-2$$ or $\eta\rightarrow A_p(\eta)=\eta'$
where $A_p$ is a lower triangle matrix with
$(1,...,1,\mu_{p,p+1}^{T_{p+1}})$ along the diagonal. It has only
the last nontrivial row except diagonal which is equal to
$(\mu_{0,p+1},...\mu_{n-2,p+1},\mu_{p,p+1})$. For simplicity we
omitted in these formulas the simplices $\sigma,T_{p+1}$. Their
presence is assumed. The full cyclic product of these matrices
gives us by definition {\bf a Nonabelian Curvature Operator}
around the $(n-2)$-simplex $\sigma$:
$$K_{\sigma,p}=A_{p+m-1}A_{p+m-2}...A_{p+1}A_p$$ of the same
algebraic form as all matrices $A_s$. Here $s$ is counted modulo
$m$. Its diagonal part is equal to
$$1,...1,\mu_{\sigma}=\prod_{s=p}^{s=p+m-1}-\mu_{s,s+1}^{T_{s+1}}=\det
K_{\sigma,p}$$ Its last row is equal to
$$\alpha_{\sigma,0,p},...,\alpha_{\sigma,n-2,p},\mu_{\sigma}$$ We
call coefficients $\alpha_{\sigma,q,p}$ and $\mu_{\sigma}$ {\bf
the Local Curvature Coefficients} where $q\in \sigma,p\in
(1,2,...,m)$. We say that our Discrete DG Connection is {\bf
Locally Unimodular, i.e. belongs to the group     $SL_n$} if $\det
K_(\sigma)=\mu(\sigma)=1$ for all $\sigma$. The Connection is {\bf
Locally Flat} if $K_{\sigma}=1$ for all $\sigma$. These
definitions imply immediately the following

\begin{lem}The local curvature operators are equal to the unit matrix if and only
if our linear equation $Q\psi=0$ has exactly $n$-dimensional space
of solutions on the universal covering space. Its  coefficients
can be computed by formula
$$\alpha_{\sigma,q,p}=\mu_{q,p}^{T_{p}}+\mu_{p-1,p}^{T_{p}}\mu_{q,p-1}^{T_{p-1}}+...+
\mu_{p-1,p}^{T_{p}}
\mu_{p-2,p-1}^{T_{p-1}}...\mu_{p-m+1,p-m+2}^{T_{p-m+2}}\mu_{q,p-m+1}^{T_{p-m+1}}$$
 In particular these coefficients transform as multiplicative
1-chains under the abelian gauge transformations
$$\psi_j=h_j\phi_j,h_j\neq 0, \psi_i\rightarrow\phi_i,
\mu_{ij}^T\rightarrow (h_i/h_j)\mu_{ij}^T$$

$$\mu_{\sigma}\rightarrow\mu_{\sigma},\alpha_{\sigma,q,p}\rightarrow
(h_q/h_p)\alpha_{\sigma,q,p}$$

\end{lem}

We can see now how to organize the simplest gauge invariant
expressions.

 \begin{lem} The quantities $$\alpha_{\sigma,q,p}\mu^{T_{p}}_{p,q}=\alpha^*_{\sigma,q,p}$$
are gauge invariant. They are connected with each other by the
formula
$$\alpha^*_{\sigma,q,p+1}=-\alpha^*_{\sigma,q,p}/(\mu^{T_p}_{pq}\mu^{T_{p+1}}_{qp})+(1-\mu_{\sigma})$$
All quantities $\alpha^*_{\sigma,q,p}$ and $\mu_{\sigma}$ can be
expressed through the gauge invariant coefficients
$\rho_{ij}^{TT'}=\mu^{T}_{ij}\mu^{T'}_{ji},T,T'\in St(\sigma)$, by
the formula
$$\alpha_{\sigma,q,p}^*=\sum_{k=0}^{k=m-1}(-1)^k\prod_{j=0}^{j=k}
\rho^{T_{p-j+1},T_{p-j}}_{p-j,q}$$

$$\prod_{p=1}^{p=m}\rho^{T_{p},T_{p+1}}_{qp}=(-1)^m\mu_{\sigma},q=0,1,...,n-2$$
so we have $n-1$ different expressions for the same quantity
$\mu_{\sigma}$.

\end{lem}

\newtheorem{cor}{Corollary}
\begin{cor}
 For the locally unimodular (i.e. locally $SL_n$) connections the
condition $\alpha_{\sigma,q,p}=0$ does not depend on the initial
point $p\in ST(\sigma)$, so the property to be locally flat in the
star $ST(\sigma)$ depends on the $n-2$-simplex $\sigma$ only.

\end{cor}
\begin{cor}
For the generic connections such that  $\alpha_{\sigma,p}\neq 0$
for all $\sigma,p$, all data $\rho_{ij}^{TT'}$ can be
reconstructed from the gauge invariant coefficients of the local
nonabelian curvature $\alpha^*_{\sigma,q,p}, \mu_{\sigma}$.

\end{cor}

Proof of the lemma. Starting with the equation expressing
$\alpha_{\sigma,q,p}$ through the collection of $ \mu$-s in the
previous lemma, we multiply both its sides by the quantity
$\mu^{T_{p}}_{p,q}$. After the elementary manipulations,  we are
coming to the expressions for $\alpha^*$ which easily implies all
statements of this lemma. Let us avoid here these elementary
calculations.

For the proof of the first corollary, we point out that the
condition $\mu_{\sigma}=1$ implies that
$$\alpha^*_{\sigma,q,p+1}=\alpha^*_{\sigma,q,p}/\rho^{T_pT_{p+1}}_{p,p+1}$$
where always $\rho\neq 0$, so our corollary obviously follows.

In order to prove second corollary, let us point out that under
the assumption of this corollary, we represent $\rho$ as a ratio
of the nonzero numbers
 $$\rho_{p,p+1}^{T_pT_{p+!}}=\alpha_{\sigma,q,p}^*/\{\alpha_{\sigma,q,p+1}^*-1+\mu_{\sigma}\}$$
This formula proves our corollary.

\vspace{0.3cm}

{\bf III.The Abelian (Framed) and Nonabelian Holonomy Groups.}

\vspace{0.3cm}

The {\bf Abelian Framed Holonomy Representation} we define for the
{\bf Framed Combinatorial Paths} starting and ending in the same
point. By definition, a {\bf Framed Combinatorial Path} is
 a sequence of  edges equipped by the $n$-simplices containing these
edges $$\gamma=<i_0,i_1,...,i_k,i_{k+1},...,i_
 N=i_0|T_0,T_1,...,T_k,...,T_{N-1}>$$ where $[i_k,i_{k+1}]\subset
T_k$. The Abelian Framed Holonomy $\mu(\gamma)$ for the framed
path $\gamma$ is equal to the product $$\mu(\gamma)=
\prod_{k}(-\mu^{T_k}_{i_k,i_{k+1}})$$ There is a natural
multiplication of the framed combinatorial paths
$\gamma_1\gamma_2$ and a whole associative semigroup of them
$\Omega=\Omega_{fr}(M,i_0)$. We have an {\bf Abelian Framed
Holonomy  Representation} $$\Omega_{fr}(M,i_0)\rightarrow k^*$$

For every framed path there is a natural {\bf Inverse Framed Path}
$\gamma^{-1}$ consisting of the same edges and $n$-simplices but
the order of passing them is reversed. The inverse framed path
leads to the inverse framed holonomy. We factorize this semigroup
by the relations $$<i,i|T>= <i,j|T><j,i|T>=1$$
$$<ij|T><jk|T><ki|T>=1$$ for the vertices $i,j,k$ belonging to the
same $n$-simplex $T$. Our relations mean exactly that such pieces
can be removed from any path if you meet these vertices one after
another as the closest neighbors. We call factor by these
relations of the semigroup $\Omega_{fr}(M,i_0)$ by the {\bf Framed
Fundamental Group}
$$\pi_1^{fr}(M,i_0)=\Omega_{fr}(M,i_0)//(Relations)$$ For the
Abelian Framed Holonomy we need only the factor-group by the
commutation relations

 $$\pi_1^{fr}(M,i_0)//(aba^{-1}b^{-1})=H_1^{fr}(M)$$
We call this factor  a  {\bf Framed Homology Group} written in the
multiplicative form.

\begin{lem}
The framed homology group is generated by the framed paths
$<i,j,i|T,T'>$ and by the arbitrary closed framed paths $\gamma_s$
whose image in the ordinary homology group (the unframed part)
generates it.

\end{lem}

\newtheorem{re}{Remark}
\begin{re}
For the choosing generators of the framed fundamental group we
have to fix for every vertex $j$ a framed path $\delta_i$ joining
the vertices $0$ and $i$. As usually, we consider the set of
closed paths $(\delta_i)<i,j,i|TT'>(\delta^{-1}_i)$ as the
additional generators in the framed fundamental group. In this
work we are dealing with the abelian case only.
\end{re}
 The Abelian Framed Holonomy leads to the representation

$$H_1^{fr}(M)\rightarrow k^*$$ The ordinary homology we obtain as
a factor-group $$H_1^{fr}//(<ij|T><ji|T'>)=H_1(M)$$ with
factorized holonomy representation $$\mu:H_1(M)\rightarrow
k^*//\{\rho_{ij}^{TT'}\}$$ for all $i,j,T,T'$.

In order to define a {\bf Full Nonabelian Holonomy Representation}
we introduce an important notion of a {\bf Thick Path} as a
sequence of the oriented $n$-simplices $\kappa=<T_1,...,T_N>$ such
that the next one is attached to the previous one along the common
$n-1$-dimensional face $\Delta$ where they induce the opposite
orientations. For the {\bf Irreducible Thick Path} the
intersections $T_k\bigcap T_{k+1}=\Delta_k$ should be exactly
equal to the $n-1$-dimensional faces for all $k=1,...,N-1$, i.e.
$T_k\neq T_{k+1}$. By definition {\bf the Closed Thick Paths with
period $N$} are defined by the condition that the last
$n-1$-dimensional out-simplex $\Delta_N$ coincides with the
initial simplex $\Delta_0$. There is also a notion of the {\bf
Periodic Thick Paths} where these sequences are infinite and
periodic. We can obviously multiply closed thick paths with the
same initial and final $n-1$-simplex $\Delta_0$. The notions of
the inverse thick path and of the trivial (empty) thick path are
natural. Therefore we are coming to the {\bf Associative Semigroup
of the Closed Thick Paths} $\Omega^{thick}(M,\Delta^{n-1}_0)$.

Let us construct a geometric model of the {\bf Abstract Thick
Paths}. We start with the standard linear $n-1$-simplex
$\Delta_0=[0,1,...,n-1]\subset R^{n-1}$ multiplied by the real
line $R$ going along the $n$-th axis $x_n$. Our abstract thick
path $\kappa_A$ will be defined by the word
$A=a_{i_q}^{r_q}...a_{i_0}^{r_0}$ of any length
$N=\sum_{s=0}^{s=q} r_s$ in the free associative semigroup with
$n$ generators $a_0,...,a_{n-1}$. As a first step, we take a
vertex $i_0$ of the initial $n-1$-simplex $\Delta_0$ for $x_n=0$
and shift it along the $n$th axis on the positive distance.  We
get new vertex $i_0'$. Now we construct a linear $n$-simplex $T_1$
with vertices $[0,...,n-1,i_0']$. It contains in the boundary an
original {\bf in-simplex $\Delta_0=[0,...,n-1]$} and a new linear
$n-1$-simplex $\Delta_1=[0,...,i'_0,...n-1]$ (the first {\bf
out-simplex}) where exactly one vertex $i_0$ is replaced by the
shifted one $i_0'$. Now taking the out-simplex $\Delta_1$ as a new
in-simplex instead of $\Delta_0$, we perform this operation once
more: we take one of the vertices of the out-simplex
$i'_1\in\Delta_1$  and shift it along the corresponding axis up on
the level higher than $i'_0$. It may be the same vertex (it should
appear exactly $r_0$ times here as in the word $A$), or another
one if we already passed all $r_0$ steps. After that we construct
a linear $n-1$-simplex $\Delta_2$ as a out-simplex for the
$n$-simplex $T_2$ and so on. Finally we are coming to the
realization of the whole word $A$ as a ''prism'' over $\Delta_0$
consisting of the $n$-simplices such that all their vertices are
located on the $x_n$-lines $R_k$ over the original vertices
$k\in\Delta^{n-1}$ with monotonically increasing heights $x_n$
except of the vertices of the  initial $\Delta_0$. This is an
abstract model of thick path with combinatorics defined by the
word $A$ consisting of the linear $n$-simplices with vertices
located in the union of the $n$ ''angle lines'' only. For every
number $k=0,1,...,n-1$ there is a subsequence of $n$-simplices in
the thick path $T_l^{(k)}\subset\kappa_A$  with shifts up along
the coordinate $x_n$ over the vertex with number $k$ in the
vertices $j_l=l\in R_k,l=0,1,2,...,N_k$ such that
$\sum_{k=0}^{k=n-1}N_k=N$, and $N_k=\sum_s r_s$ where $i_s=k$. We
call these sequences
$<j_0,j_1,...,j_{N_k}|T_0^{(k)},...,T_{N_k}^{(k)}>$ {\bf the
abstract angle framed paths}.

 Any thick path can be realized  in the manifold $M$ with the
initial oriented $n-1$-simplex  $\Delta_0$ is given. An initial
oriented simplex $\Delta_0$ uniquely determine the irreducible
thich path with given combinatorics. All $n$-simplices $T_k$
should be attached to the previous oriented $n-1$ out-simplex
$\Delta_{k-1}$ inducing in it the right orientation, and $T_k\neq
T_{k-1}$  in the irreducible case. For the realization of the
reducible thick path we should indicate the ''turning points'' in
the sequence of simplices. An irreducible  thick path will be
determined by the combinatorics of the word $A$ only. Topology of
the triangulated manifold $M$ determines which paths with the
initial face $\Delta=\Delta_0$ are in fact closed. {\bf The set of
closed thick paths started in the face $\Delta_0$ we denote by
$\Omega^{thick}(M,\Delta_0$)}. For the closed thick path the
corresponding angle framed paths will not necessarily be closed in
the manifold $M$:
\begin{lem} There is a natural homomorphism into the permutation group $S_n$
$$P:\Omega^{thick}(M,\Delta_0)
\rightarrow S_n $$ induced by the permutation of the vertices of
the $n-1$-simplex  after the identification of the initial
in-simplex $\Delta_0$ and the last out-simplex in the closed thick
path.
\end{lem}
The kernel of this representation will be denoted $\ker P=
\Omega^{thick}_0(M,\Delta_0)$. Let a closed   thick path
$\kappa\in\Omega^{thick}(M,\Delta_0)$ in the manifold $M$ be given
starting and ending in the $n-1$-face $\Delta_0$, and let
$0,1,...,n-1$ be its vertices. For any Discrete DG connection with
coefficients $\mu_{ij}^T$ we define a {\bf Nonabelian Holonomy
Representation} along the closed thick path $\kappa$ with
permutation $P_{\kappa}$:

 Starting from the initial data $\eta=(\psi_0,...,\psi_{n-1})\in R^n_{\Delta_0}$,
 we step by step calculate
the values of function $\psi$ in the vertices of the thick path
$\kappa$ from the equation $Q\psi=0$. For the thick path
$\kappa=(T_N\ldots T_1)$ it leads to the linear map
$$\tilde{K}_{\kappa}:R^n_{\Delta_0}\rightarrow
R^n_{\Delta_N}=\tilde{K}_{T_N}\ldots\tilde{K}_{T_1}$$ where
$$\tilde{K}_T:R^n_{\Delta}\rightarrow R^n_{\Delta'}$$ is the
one-step map from the {\bf in-face} into the {\bf out-face} for
the $n$-simplex $T$ provided by the DG connection.

 By the {\bf
Nonlinear Holonomy Map} we call a product $$
K_{\kappa}=P_{\kappa}\tilde{K}_{\kappa}$$ So the correspondence
$\kappa\rightarrow K_{\kappa}$ generates a holonomy representation
$$K:\Omega^{thick}(M,\Delta_0)\rightarrow GL_n(k)$$
\begin{lem}The Holonomy Representation is a Homomorphism
$$\Omega^{thick}(M,\Delta_0)\rightarrow GL_n(k)$$
$$K_{\kappa_2}K_{\kappa_1}=K_{\kappa_2}K_{\kappa_1}$$
\end{lem}
For the proof of this lemma  we point out that $K=P\tilde{K}$.
From the definition of $\tilde{K}$ we have:
$$(P_{\kappa_1}^{-1}\tilde{K}_{\kappa_2}P_{\kappa_1})\tilde{K}_{\kappa_1}=\tilde{K}_{\kappa}$$
because the basis of vertices of the $n-1$-simplex $\Delta_0$ is
shifted by the permutation $P_{\kappa_1}$ after passing the first
closed path $\kappa_1$. Therefore we obtain finally
$$K_{\kappa_2\kappa_1}=(P_{\kappa_2}P_{\kappa_1})P_{\kappa_1}^{-1}\tilde{K}_{\kappa_2}P_{\kappa_1}
\tilde{K}_{\kappa_1}=K_{\kappa_2}K_{\kappa_1}$$ Lemma is proved.

For the unique nontrivial closed thick path in the simplicial star
$\kappa\subset ST(\sigma)$ surrounding $n-2$-dimensional simplex
$\sigma$, this holonomy map reduces to the ''Nonabelian Curvature
Map'' $K_{\sigma,p}$ defined in the
 previous paragraph where $\Delta=[\sigma,p]$. This simplest path
corresponds to the most elementary word $A=a_j^m$ rotating
$n$-simplices around the $n-2$-simplex $\sigma$ opposite to the
vertex $j\in\Delta$.

\begin{lem}
For every closed thick path $\kappa_A$ determined by the word $A$
and initial $n-1$-simplex $\Delta_0$ the determinant of the
Nonabelian Holonomy Map has a form $$\det
K_{\Delta_0}=(-1)^N\prod_{k=0}^{k=n-1}\mu_{k}(\kappa_A)$$ where
$$\mu_k(\kappa_A)=\prod_{l=0}^{l=N_k}- \mu^{T_l^k}_{i_l,i_{l+1}}$$
is the product along the ''angle'' axis $R_k$ going up with the
variable $x_n$ from the vertex corresponding to $k\in\Delta_0$ in
the abstract model of the thick path. The quantities
$\mu_{k}(\kappa_A)$ are equal to the Abelian Framed Holonomy
Representation of the  framed paths
$$\gamma_{A,k}=<j_0,...,j_{N_k}|T^k_0,...,T^k_{N_{k-1}}>$$ called
  the ''angle'' paths of the thick path $\kappa_A$
$$\mu_k(\kappa_A)=\mu(\gamma_{A,k})$$ The product of all angle
paths is closed.

If all local curvature operators $K_{\sigma,p}$ are equal to the
unit matrix for all $n-2$-simplices $\sigma$, then the Nonabelian
Holonomy depends on the class of thick path in the fundamental
group $\pi_1(M)$ only.

\end{lem}

\begin{lem}
All matrix elements $\alpha_{ij}(\kappa,\Delta_0)$ of the
operators $K=P\tilde{K}$ of the Nonabelian Holonomy transform
under the abelian gauge transformations as one-dimensional
multiplicative cochains $\alpha_{ij}\rightarrow
h_i/h_j\alpha_{ij}$ where $i,j$ are the vertices of the initial
$n-1$-simplex $\Delta_0$ where $$K\rightarrow
HKH^{-1},H=diag(h_1,...,h_{n-1})$$
 For the local
Nonabelian Curvature Operators we have $P=1,\Delta_0=[\sigma,p]$
where $\sigma$ is an arbitrary $n-2$-simplex, and $p=1,...,m$ is a
vertex in its simplicial star $p\in ST(\sigma)$. All
gauge-invariant polinomial in the variables
$\alpha_{ij}(\kappa_A),\mu_{ij}^{TT'}$ can be expressed as
polinomials from the  the framed abelian holonomy of the closed
paths in $M$.
\end{lem}

Proof of this lemmas presents no difficulties. For the proof let
us point out that our matrix elements $\alpha_{ij}$ can be
expressed as the sums of the products  of the quantities
$\mu_{i_l,j_l}^{T_l}$ along the paths easily visible as the paths
monotonically going up (see Fig) in the abstract model of thick
path, starting in the simplex $\Delta_0$ and ending in the upper
$n-1$-simplex (who coincides with $\Delta_0$ for the closed paths
in the manifold.) Therefore after the gauge transformations only
the boundaries of the paths will give contribution, so only the
ends $i,j$ remain in the final answer. At the same time, all gauge
invariant expressions presented as polynomials of the path
integrals of the quantities $\mu$ can be expressed through the
holonomy of the closed paths (no free ends can be left for the
gauge invariant expression). Our lemma is proved.

\begin{ex} Let us consider an interesting example of the {\bf Canonical
Connection} on the triangulated manifolds $M$ where all connection
coefficients are equal to one in every simplex $b_{T:i} =1$. We
have also $\mu_{ij}^T=-1$. This connection has been considered in
\cite{ND3}. It appeared also in the work \cite{J} in the different
terminology for the needs of the coloring problem. Its image
belongs to the group $S_{n+1}$ but we normally realize this group
linearly
 $S_{n+1}\subset GL_n$ using the imbedding $R^n\subset R^{n+1}$
as a subspace invariant under the permutation of all coordinates
$\psi_i$ where $\sum\psi_i=0,i=0,...,n-1$. Exactly that
corresponds to the canonical connection in the work \cite{ND3}.
Starting from any $n-2$-simplex $\Delta_0$, we construct a {\bf
Coloring of the Vertices by the $n+1$ colors $u_0,...,u_n$} along
the thick path with combinatorics corresponding to the word $A$.
Assigning to the initial vertices of the in-simplex $\Delta_0$ the
colors $u_0,...,u_{n-1}$, we can see that the final coloring of
the vertices of the out-simplex is uniquely defined by two
factors:
 by the
combinatorics of the word $A$ and by the realization of this final
out-simplex in the manifold $M$, i.e. by the permutation
$P(\kappa)$. We denote a nonabelian holonomy map associated with
this connection by the
$$K_{\kappa}^{can}=P_{\kappa}\tilde{K}^{can}_{\kappa}$$

\begin{lem}
For any closed thick path $\kappa$ with combinatorics
corresponding to the word $A=a_{i_q}^{r_q}...a_{i_0}^{r_0}$ the
resulting permutation corresponding to the holonomy of the
canonical connection is given by the formula
$$K^{can}_{\kappa}=P_{\kappa}\tau_{i_q,n}^{r_q}...\tau_{i_0,n
}^{r_0}$$ where $i=0,1,...,n-1$ and $\tau_{i,n}$ is a permutation
of the points $i,n$ only, $\tau^2_{i,n}=1$. The permutation $P$ in
this formula permutes only the numbers $0,1,...,n-1$ leaving the
index $n$ invariant.

\end{lem}
Proof of this lemma follows immediately from the definition of the
abstract model of the thick path.

\end{ex}

\vspace{0.3cm}

{\bf III. Solution of the Reconstruction Problem for the case
$n=2$}

{\bf Flat connections. The case $n\geq 2$}.

\vspace{0.2cm}

Consider now any oriented trangulated 2-manifold $M$ with the
vertices $i,j,...$, 2-simplices $T,T',...$ and with the discrete
DG Connection. Our field is $k=R,C$ only.

\begin{th}
All coefficients $\mu_{ij}^T$ of the discrete DG Connection over
the field $k$  can be recovered up to  abelian gauge
transformation from the abelian framed holonomy representation
$$\mu:H_1^{fr}(M)\rightarrow k^*$$ where the framed abelian
holonomy image of the group $H_1^{fr}(M)$ is generated by the
elements $\rho_{ij}=\mu_{ij}^T\mu_{ji}^{T'}=<i,j,i|TT'>^*$ and by
the images of the generators of the ordinary homology group
$\mu(\gamma_s),\mu(a_q)\in k^*$.

\end{th}

Proof. Let us describe the reconstruction process. Introduce the
new quantities by the following formula:
$$\lambda_{ij}=-\mu^{T}_{ij}/\sqrt{\rho_{ij}^{TT'}},\partial
T=[ij]+...,\partial T'=[ji]+...$$ where some specific value of the
square root is chosen. If our manifold is oriented,  we choose the
orientation of the triangles $T,T'$ to be the same as the global
orientation of manifold, and $\partial T=[ij]+...$. In that case
we forget about the indices $TT'$ in the formulas, so we have
$\lambda_{ij}=-\mu_{ij}/\sqrt{\rho_{ij}}=\sqrt{-\mu^T_{ij}/\mu^{T'}_{ji}}$
where $$\rho_{ij}=\rho_{ji}, \lambda_{ij}=\lambda_{ji}^{-1}$$
Nonuniqueness of choosing the square root we resolve by choosing
square roots separately defining $\sqrt{-\mu_{ij}^T}$ in every
triangle $T$ with requirement
$$\sqrt{-\mu_{ij}}\sqrt{-\mu_{jl}}\sqrt{-\mu_{li}}=1$$. We shall
return to these details later.

\begin{lem}
For the coboundary of the $k^*$-valued multiplicative cochain
$\lambda=(\lambda_{ij})$ defined by the formula $$d\lambda
(T)=\lambda(\partial T)=\lambda_{ij}\lambda_{jl}\lambda_{li},
T=[ijl]$$ we have $$d\lambda
(T)=(\rho_{ij}\rho_{jl}\rho_{li})^{-1/2}=\rho^{-1/2}(T)$$
Therefore for every finite triangulated domain $D$ in the manifold
$M$ following integral formula is true expressing the integral of
the ''Curvature'' $(\rho (T))^{-1/2}$ along the domain $D$ through
the framed abelian holonomy of the boundary curves $\partial
D=\bigcup_q\gamma_q$ with framing by the triangles looking in the
external direction to the domain $D$: $$\prod_{T\in
D}(\mu_{ij}^T)^{-1/2} =(\mu(\partial D))^{-1/2}$$ In particular,
for the compact oriented manifold $M$ we have $$\prod_{T\in M}
(\rho (T))^{-1/2}=1$$

\end{lem}
The proof of this lemma follows immediately from  definition of
the quantities $\rho_{ij}$ and $\lambda_{ij}$ taking into account
the equality $\mu_{ij}\mu_{jl}\mu_{li}=-1$ and our agreement that
$\mu_{ij}^T=\mu_{ij}$ for the right orientation leading to the
conclusion that $\mu_{ji}=\mu^{T'}_{ji}$ where $T'\neq T$, and
$\rho_{ij}=\mu_{ij}\mu_{ji}=\rho_{ji}$. Our Lemma is proved.

As a corollary, we are coming to the following conclusion: knowing
$\rho_{ij}$ we can reconstruct an unknown cochain $\lambda$. After
that we define our DG Connection by the formula
$$\mu_{ij}=-\lambda_{ij}\sqrt{\rho_{ij}}$$

By definition, this is a solution of our system. A cochain
$\lambda$ is nonunique: any cocycle $\delta$ may be used to change
it: $\lambda'=\lambda\delta$, i.e.
$\lambda_{ij}'=\lambda_{ij}\delta_{ij}$ where
$\delta_{ij}\delta_{jl}\delta_{li}=1$ for ever triangle $[ijl]$.
It is obvious that there is nothing except the set of all
1-cocycles $\delta$ and all possible changes of signs in the
definition of the square roots of $\rho_{ij}$ what may lead to the
same set of the data $\{\rho_{ij}\}$. Making an arbitrary abelian
Gauge transformations $\mu_{ij}\rightarrow
\mu_{ij}'=(h_i/h_j)\mu_{ij}$ we change $\lambda$ by the cocycle
$\delta_{ij}=h_i/h_j$, i.e. by the coboundary. Changing signs of
the square roots  $$(\rho_{ij})^{1/2}\rightarrow -
(\rho_{ij})^{1/2} $$ we  change $\lambda$ by the same signs, so
the resulting value of $\mu$ remains unchanged.

Changing $\lambda$ by the cocycle $\delta$ nonhomologous to zero,
we also change $\mu$ by the same $\delta$, i.e.
$\mu\rightarrow\mu\delta=\mu'$. Therefore our framed abelian
holonomy along the closed contours will be changed by the
integrals
$$\mu(\gamma)\rightarrow\mu(\gamma)\prod_{[ij]\in\gamma}\delta_{ij}$$
Now we are fixing $\tau$ by the requirement of the theorem that
the framed abelian holonomy is prescribed along some basis
$\gamma_k$ of the homology group $H_1(M)$. Our theorem is proved.

\begin{lem}
Let $\lambda_{ij}^{TT'}=\mu^T_{ij}/\sqrt{\rho_{ij}^{TT'}}$ and
$\lambda_{ij}^{TT'}\rightarrow\lambda_{ij}^{TT'}\delta_{ij}$,
$\mu_{ij}^{T}\rightarrow \mu_{ij}^T\delta_{ij}$ where $\delta$ is
an ordinary multiplicative cocycle. Then the framed abelian
holonomy is changed by the '' multiplicative integral'' of
$\delta$ along the same closed paths. The operators of Nonabelian
Holonomy along the Thick Paths are changed in the following way:

$$K_{\kappa}\rightarrow C^{-1}HK_{\kappa}H^{-1}$$ where
$C=\delta_{\kappa}$ is a value of the 1-cocycle $\delta$ on the
element $\kappa\in \pi_1(M)$ of the fundamental group, $\Delta_0$
is an initial $n-1$-simplex of the thick path $\kappa$ and
$H=diag(\hat{h}_0,...,\hat{h}_{n-1})$ is the diagonal matrix whose
entries are well-defined up to the common nonzero multiplier, and
$\hat{h}_i/\hat{h}_j=\delta_{ij}$.
\end{lem}

Proof. The framed abelian part of this lemma is obvious from
definitions of $\lambda$ and $\delta$. In order to prove
nonabelian part, we consider this Discrete DG Connection on the
special abelian covering $\pi:\hat{M}\rightarrow M$ such that our
cocycle became exact $\pi^*\delta=d\hat{h}$. Consider any closed
thick path $\kappa$ starting and ending in the $n-1$-simplex
$\Delta_0$ of the manifold $M$ and realizing a generator
$\gamma\in H_1(M)$. Without any losses of generality, we may think
that our cocycle $\delta$ has nontrivial ''multiplicative
integral'' along this basic element only, and that our covering is
cyclic. On the covering space $\hat{M}$ we choose a covering
$n-1$-simplex $\Delta,\pi(\Delta)=\Delta_0$. After that we get a
unique covering thick path $\hat{\kappa}$ starting  in $\Delta$.
This path is a covering path over the closed path $\kappa$ with
period $N$, i.e. it consists of the  sequence of $n$-simplices
$...,T_1,...,T_{N-1},T_N,...$ such that $\pi(T_i)=\pi(T_{N+i})$,
and $T_{N-1}\bigcap T_N\in (\pi)^{-1}(\Delta_0)$. The monodromy
map on the covering space
$R=\hat{\gamma}:\hat{M}\rightarrow\hat{M}$ transforms thick
covering path into itself, and $R(T_1)=T_N,
R(\Delta)=T_{N-1}\bigcap T_N$. Consider the function $\hat{h}$ in
the covering thick path. It is nonperiodic: we have
$R^*(\hat{h})=C\hat{h}$ where $C=\prod_{[ij]\in\gamma}\delta_{ij}$
by definition. In the covering path our DG Connections both are
periodic and gauge (abelian) equivalent to each  other but the
equivalence is nonperiodic. According to the lemma 6 (above) we
can see that the matrix elements of our Nonabelian Holonomy
Operator  transform by the following formula
$$\alpha_{ij}\rightarrow \hat{h}_i/R^*(\hat{h}_j)\alpha_{ij},
i,j\in\Delta_0$$ because our thick path starts at $\Delta$ and
ends at $R(\Delta)$ in the covering space. At the same time, we
have $R^*(\hat{h})=C \hat{h}$.This is exactly the statement of our
lemma. Lemma is proved.

\begin{th}
For every data $\rho_{ij}^{TT'}$ on the orientable 2-manifold $M$
corresponding to flat $GL_2$ connections (i.e. with trivial
nonabelian local curvature) , there exist exactly one
$SL_2$-connection up to abelian gauge transformation and changing
sign.

\end{th}
\begin{re}The existence of $SL_n$-connection follows from the same arguments
 also for all $n\geq 2$,
but the uniqueness  for $n>2$ will be proved later using some
additional arguments not presented yet(see below).
\end{re}

 Proof of this theorem
follows from the previous lemma: we may change determinant on the
Nonabelian Holonomy Group multiplying $\rho_{ij}$ by the 1-cocycle
$\delta_{ij}$. Let us point out that after making determinant
equal to 1, we may also change sign of the holonomy.

 {\bf The first Chern Number:} Let now
$n=2$. Consider the case $k=C$ and assume that
$|\arg[\rho_{ij}^{TT'}]|<\pi/2$. We define an integer-valued
cohomology class $c_1\in H^2(M,Z)$ by the formula for the cochain:
$$c_1(T)=\frac{1}{2\pi i} \arg [(\rho(T))^{-1/2}]$$ From the
equality $$\prod_T (\rho(T))^{-1/2}=1$$ for the closed oriented
manifold we obtain the result $$\sum_{T\in M} c_1(T)=r\in Z$$

Our condition permits us to make such unique  choice that $$|\arg
[\rho (T)^{-1/2}]|< \pi$$ in all cases. With this agreement we are
coming to the well-defined integer number.

 \vspace{0.3cm}

{\bf IV.Multidimensional Discrete DG Connections. }

\vspace{0.3cm}

Consider now any $n>2$. We  expect that all DG Connection can be
reconstructed from the framed abelian representation. Let a closed
oriented triangulated manifold $M$ be given, and $s_k$ are the
numbers of $k$-simplices. We know a few number of general
relations for these numbers: $$s_{n-1}=\frac{n+1}{2}s_n$$
$$\sum_{k=0}^{k=n}(-1)^ks_k=\eta(M)$$ where $\eta(M)$ is the Euler
characteristics. In particular, $\eta=0$ for the odd dimensions
$n=2t+1$. Let us present the numbers $s_k$ in the form:
$$s_k=\frac{(n+1)!}{(k+1)!(n-k)!m_k}s_n $$ The meaning of this is
following: every $n$-simplex has exactly $(n+1)!/(k+1)!(n-k)!$
faces of the dimension $k$. If every $k$-simplex belongs exactly
to $m_k$ $n$-dimensional simplices, one can deduce that this
number $s_k$ can be computed exactly as it is written here. For
example, we have always $m_{n-1}=2$ in the manifolds, $3\leq
m_{n-2}<m_{n-3}<...<m_0$. In general, this formula gives
definition of the numbers $m_k$ as some sort of {\bf the mean
value of the number of $n$-simplices containing a random
$k$-simplex}.

\begin{ex} As a simplest example we take sphere $M=S^n=\partial
\Delta_{n+1}$ as a boundary of the $(n+1)$-simplex. In this case
$m_{n-k}=k+1$.
\end{ex}
 Let us try to count a number of parameters in our reconstruction
problem.
\begin{lem} The number $(\mu)$ of  independent quantities
 $\mu^{TT'}_{ij}$ modulo the  abelian gauge transformations
is equal to $ns_n-s_0+1=(\mu)$.
\end{lem}

 Proof. In every $n$-simplex
$T=[0,1,...,n]$ we have exactly $n$-dimensional manifold of the
quantities $\mu_{ij}^T$ with (multiplicative) basis $\mu_{0i}^T$
according to the relations  indicated in their definition above.
Different $n$-simplices are completely independent. Applying the
abelian gauge transformations we extract exactly $s_0-1$ parameter
because the constant function leads to the trivial gauge
transformation. This argument implies the statement of our lemma.

  For every $n\geq 2$ we define the number  $(\rho)$ equal to the
(multiplicative) dimension of the set of all quantities
$\rho_{ij}^{TT'}$ plus $b=b_{k^*}$ where $b$ is the rank of the
Betti number $b_1$ for the real positive holonomy representation
$k^*=R_+$, and it is equal to the larger number
$b^*_1=b_1+torsion_1$ for $k^*=C^*$. We present it by the
expression $$(\rho)=(n-1)s_{n-1}-(n-2)s_{n-2}+b+R$$
 where $R$ is the  remaining part.

 In order to explain  meaning of this phrase,  let us point
 out that by the
same reason there exist exactly $n-1$-dimensional  space of
quantities $\rho_{ij}^{TT'}$ in every $n-1$-simplex $T\bigcap T'$.
However, they are not independent for different pairs of neighbors
$T,T'$. According to the lemmas above (see local nonabelian
curvature) there are $n-1$ different expressions for the quantity
$\mu(\sigma)=\det K_{\sigma,p}$ in the star $St(\sigma)$ of every
$n-2$-simplex $\sigma$ depending on the vertex of this simplex
$\sigma$. It is very probable that the number of remaining
relations does not depend on triangulation (at least for $n=3$.
What is important is that these relations are not
 independent for $n>2$ in general as we shall see below. Therefore some unknown number
$R$ enters  our calculation making the answer totally unclear.

\vspace{0.3cm}

\begin{ex}

 Consider the simplest case $n=2$ where everything is already
known. The number $(\mu)$ is described by this formula for all
$n\geq 2$. For the number $(\rho)$ we have $(\rho)=s_1-1+b$ where
$b=2g$ and $R=-1$ because there exist a nontrivial global relation
$\prod_{ij,T,T'\in M}\rho_{ij}^{TT'}=1$. Taking into account the
relations $s_2=3/2s_3,\eta(M)=s_2-s_1+s_0=2-2g$, we are coming to
the equality $$(\mu)=(\rho)$$ for the closed oriented 2-manifolds.
The neighboring oriented pairs $TT'$ are chosen such that
$\partial T=[ij]+...,\partial T'=[ji]+...$. It corresponds to the
fact established above that for the reconstruction of the DG
Connection we have to include in the data also values of the
framed abelian holonomy on the set of $2g$ closed paths presenting
the basis of the $H_1(M)$ except all $\rho_{ij}$.
\end{ex}

\vspace{0.3cm}

\centerline{\bf The case $n\geq 3$}

\vspace{0.3cm}

\begin{ex} For the case $n=3$ we have $(\mu)=3s_3-s_0+1$ and $s_2=2s_3,
s_3-s_2+s_1-s_0=0$. Therefore we obtain as a corollary of the
relation $3s_3-s_0+1=2s_2-s_1+b+R$ that $$R+b=1$$ For the
homological 3-spheres we have $b=0$ and $R=1$.
\end{ex}

 We are going to
prove below that the framed abelian holonomy uniquely determines
the DG Connection  up to the abelian gauge transformation for all
$n\geq 3$. Therefore  the dimension $\rho^{realizable}$ generated
by all $\rho_{ij}^{TT'}$ generated by the DG Connections is equal
for $n\geq 3$ to the number $(\mu)-b$. For example, we have for
$n=3$: $$ \rho^{realizable}=3s_3-s_0+1-b= 2s_2-s_1+R$$ where
$R=1-b$. The space of all functions with formal properties of the
quantities like $\rho_{ij}^{TT'}$ and general local relations
indicated above for $n=3$ has dimension equal to $2s_2-s_1$ plus
something depending on the topology of the manifold $M$ only.
Looking on the right-hand side of this relation, we expect to find
out {\bf exactly one global dependence} between the $s_1$ already
known general ''local'' relations for these quantities in the
simplicial stars of all $n-2$-simplices, {\bf plus exactly $b$
''global'' relations} depending on the 1-cycles in the
$n$-manifold $M$ as a minimal possibility. We shall describe these
relations below for the closed oriented 3-manifolds. Let now
$n\geq 3$.

\begin{th}
Let $k^*=C^*$.

I. For every integral 2-cycle $z=\sum \Delta_k,a_j\in Z$ presented
as a sum of the oriented 2-simplices in the n -manifold $M$ there
is a relation between the quantities $\rho_{ij}^{TT'}\in C^*$: Let
an arbitrary ''framing'' be chosen along this 2-cycle, i.e. with
every 2-simplex $\Delta_k\subset z$ we associate an n-simplex $T$
such that $\partial T=\Delta_k+...$. This relation has a form
$$\prod_{[ij]\in \Delta_k\bigcap\Delta_{k'}\subset
z}\rho_{ij}^{T_kT_{k'}}=1$$

II.For every 1-dimensional ''torsion cycle'' $a\in H_1(M,Z)$ of
the order $r$ (i.e. $a^r=1$ in the multiplicative form)
 there is a relation between the abelian framed holonomy
$\mu(a)$ along the framed path $a$ and the quantities
$\rho_{ij}^{TT'}$.  Let an integral chain (simply, a formal sum of
the oriented 2-simplices) $u=\sum_k \Delta_k$ with some framing
$T_k,
\partial T_k=\Delta_k+...$ is given such that $\partial
u=a^r$. The relation has a form $$\prod_{[ij]\in \Delta_k\bigcap
\Delta_{k'}\subset u}\rho_{ij}^{T_kT_{k'}}=\mu(\partial
u)=\mu(a)^r$$, and $(\partial u)$ is a boundary 1-chain with
induced framing. The product  along the chain $u$ is defined
representing our chain $u$ through the pairs of 2-simplices
$\Delta_k,\Delta_{k'}$ whose intersection is exactly the edge
$[ij]$ entering them with the opposite orientations.

\end{th}

\begin{cor}
For the closed oriented 3-manifolds $M$ we have $b_2=b_1$ by the
Poincare duality. Therefore we have $b=b_2+torsion_1$, and the
number of relations is equal to the rank $b$ of the topological
part of the framed abelian holonomy.
\end{cor}

The proof of this theorem follows from the integral formula:

\begin{lem}
For every integral 2-chain $w=\sum_k\Delta_k$ with given framing
of the 2-simplices following ''Integral Formula'' is true:
$$\prod_{[ij]\in \Delta_k\bigcap\Delta_{k'}\subset
w}\rho_{ij}^{TT'}=\prod_{[ij ]\in\partial w}-\mu_{ij}^T$$
\end{lem}

\centerline{\bf The first Chern class}.

As a by-product of this theorem we define a first Chern class
$$c_1\in H^2(M,Z)$$ in the same way as for $n=2$ above: we put
$$(c_1,z)=(2\pi)^{-1}\sum_{[ij]\in \Delta_k\bigcap \Delta_{k'}}
\arg [\rho^{T_kT_{k'}}_{ij}]\in Z$$ for every cycle $z$; For the
cycles modulo finite order we define $$(c_1,u)=(2\pi)^{-1}\sum
\arg \rho_{ij}^{T_kT_{k'}}\in Z/rZ$$ as in the theorem above. The
topological properties of these quantities in the discrete case
should be especially discussed. We avoid this discussion here.

{\Large Reconstruction of the Connection for $n\geq3$}

Let us start with the case $n=3$. Consider now a three-dimensional
oriented manifold $M$. For every simlicial star $St(\sigma)$ of
one dimensional simplex $\sigma=(ij)\subset M$ we reconstruct the
quantities $\mu_{ij}^T$ up to the unknown constant $\delta_{ij},
T\in St_{\sigma}$, from the equations
$$\rho^{T_{p}T_{p-1}}_{ij}=\mu^{T_{p}}_{ij}\mu^{T_{p-1}}_{ji},p=1,2,\ldots
q(\sigma) $$ These equations are solvable because the trivial set
of  relations B is satisfied (see Lemma 1). Here the simplices
$T_p\in St(\sigma)$ are numerated in the natural cyclic order
modulo $q(\sigma)$ where $T_{q+1}=T_1$. So we know the quantities
$$\tilde{\mu}_{ij}^{T_p}=\mu_{ij}^{T_p}\delta_{ij}$$ where
$\delta_{ij}\delta_{ji}=1$, if solution
$\mu_{ij}^T=(\mu_{ji}^T)^{-1}$ exists. Let us consider the
necessary equation $$\mu_{ij}^T\mu_{jl}^T\mu_{li}^T=-1$$ We are
coming to the conclusion that our problem is solvable if and only
if following three requirements are satisfied: 1.The quantity
$$\tilde{\mu}_{ij}^T\tilde{\mu}_{jl}^T\tilde{\mu}_{li}^T=\tilde{\mu}[\Delta]^T$$
is in fact cochain depending only on the oriented simplex
$[ijl]=\Delta$, i.e. $\tilde{\mu}[ijl]\tilde{\mu}[jil]]=1$; 2.
This cochain is closed; 3.This cochain is exact.

Proof of the Statement 1: We define this quantity on the oriented
manifolds $M$ for the 3-simplices $T$ of the right orientation
such that $\partial T=[ijl]+...$. This agreement makes it
well-defined as a function of oriented simplices. In order to
prove that $\tilde{\mu}[ijl]=\tilde{\mu}[jil]^{-1}$, we use the
identity $$\rho_{ij}^{TT'}\rho_{ jl}^{TT'}\rho_{li}^{TT'}=1$$ for
the pair of oriented simplices such that $\Delta=T\bigcap T'$. We
know that
$\rho_{ij}^{TT'}=\tilde{\mu}^T_{ij}\tilde{\mu}^{T'}_{ji}$. This
equality immediately implies our result. The statement 1 is
proved.

Proof of the Statement 2.  For every 3-simplex $T$ we consider a
full product of the quantities $\tilde{\mu}[\Delta_s]$ along the
boundary simplices $\Delta_s,s=1,2,3,4$. It is easy to see that
for every edge $ij\subset T$ we have exactly two multipliers in
this product equal to $\tilde{\mu}^T_{ij}$ and
$\tilde{\mu}^T_{ji}$. We use here the result of the statement 1
expressing everything through the quantities $\tilde{\mu}^T$ for
the faces of any orientations (other 3-simplices are not needed).
This property implies our result. The statement 2 is proved.

Proof of the Statement 3. We know already that this is a
multiplicative cocycle with values in $C^*$. As everybody knows,
its exactness requires the conditions formulated in the Theorem
above where the homological relations for the connection
coefficients were found. Therefore the statement 3 follows from
the homological arguments.

Using this result, we easily reconstruct our connection
$\mu_{ij}^T$: we take any solution to the equation
$$d\delta=\tilde{\mu}[\Delta]$$ and put
$\mu^T_{ij}=\delta_{ji}\tilde{\mu}^T_{ij}$. This solution
satisfies to all requirements and define exactly the same local
part of the framed abelian holonomy representation. This set of
quantities can be choose modulo closed 1-cocycle
$\delta_{ij}\rightarrow \delta'_{ij}$ such that their ratio is
closed. This degree of freedom should be used in order to make the
proper adjustment of the global part of  framed abelian holonomy
along the basis of the first homology group. The exact part of
this cocycle is responsible for the abelian gauge transformations.
These arguments finish our problem.

Nonorientable case can be easily reduced to the oriented one using
the orientable 2-covering in the same way as for the case $n=2$.

For the manifolds $M$ of all dimensions $n\geq 3$ we develop the
same reconstruction process as for $n=3$.

Step 1. Consider the simplicial star $St(ij)$ of the edge $ij$.
Solve the equations
$$\rho_{ij}^{TT'}=\tilde{\mu}^T_{ij}\tilde{\mu}^{T'}_{ji}$$ in the
star. This problem can be solved uniquely up to unknown constant
$\delta$: $$\mu^T_{ij}=\delta_{ij}\tilde{\mu}^T_{ij},
\delta_{ij}\delta_{ji}=1$$ This is true because $\rho^{TT'}_{ij}$
is a 1-cocycle in the dual cell decomposition of the star $St(ij)$
where $T$ are the vertices (see the relations  B in Lemma 1). So
our condition for the solvability of that intermediate problem is
$H^1(St(ij),k^*)=1$. It is certainly true in the manifolds.

Step 2. In order to solve the equation
$\mu^T_{ij}\mu^T_{jl}\mu^T_{li}=-1$ for every $n$-simplex $T$ we
need to prove that this quantity
$$\tilde{\mu}^T_{ij}\tilde{\mu}^T_{jl}\tilde{\mu}^T_{li}=\tilde{\mu}^T[ijl]$$
is in fact a well-defined multiplicative cocycle independent on
$T$. This statement follows from the requirement
$$\rho^{TT'}_{ij}\rho^{TT'}_{jl}\rho^{TT'}_{li}=1$$ where
$\rho^{TT'}_{ij}=\tilde{\mu}^T_{ij}\tilde{\mu}^{T'}_{ji}$. So we
conclude that $\tilde{\mu}^T[ijl]\tilde{\mu}^{T'}[jil]=1$ for
every pair $T,T'$. The proof that this 2-cochain is closed is the
same as for the case $n=3$.

Step 3. In order to prove that this cochain is exact we use an
analog of the same relations for the quantities $\rho$ integrated
along the cycles (see the Theorem above). This theorem gives us
the  set of relations which leads to the property of any cocycle
to be exact in the elementary homological algebra.

After that all arguments  coincide with the case $n=3$. Our
reconstruction process is finished.

In particular, we see that the Uniqueness Theorem for all $n\geq
2$ follows from our results:

\begin{th}
The    framed abelian holonomy representation determines
completely the Discrete $GL_n$-Connection $\{\mu_{ij}^T\}$ on the
triangulated  n-manifold $M$ up to the abelian gauge
transformation. The set of conditions on the data of the framed
abelian holonomy representation found in this work is necessary
and sufficient for the reconstruction.

\end{th}

\end{document}